# Viewpoint | Krzysztof R. Apt

# One More Revolution to Make: Free Scientific Publishing

Computer scientists are in the position to create new, free high-quality journals. So what would it take?

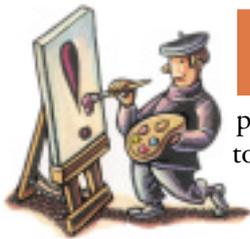

LISA HANEY

**D**eclining costs of access to information have been a crucial factor in the progress of humanity. Thanks to the Internet it is now feasible to provide and properly organize a freely available scientific knowledge. It is the scientists' responsibility to work toward this goal. We can achieve it only by changing our attitudes toward scientific publishing. Computer scientists are in a position to create new, free high-quality journals and contribute in this way to a free dissemination of scientific knowledge. We should draw our inspiration from the success of the GNU project (www.gnu.org), started by Richard Stallman and joined by leading programmers, the aim of which is to develop free software as a viable alternative to commercially produced software. We should also listen to mathematicians and economists [1, 2] who convincingly argue for the creation of new inexpensive mathematics and economics journals that would supersede the over-priced ones.

## Anomalies of Scientific Publishing

You don't need to be a genius to realize that scientific publishing is a very peculiar branch of business. Here is a list of some of its striking aspects:

• The prices for comparable-in-quality products can differ drastically. Take, for example, two similiar-in-scope journals: *ACM Transactions on Programming Languages and Systems* and Elsevier's *Science of Computer Programming* (SCP). In 1999 the ratio in their prices per page for printed versions was close to 1:7. (Incidentally, in 1982, just after SCP was created, this ratio was 1:4.) Peculiarities exist in other fields of science (see [1] for an assessment of the situation in the field of economics).

• Scientific publishing it is the only branch of industry that relies on massive voluntary work. This voluntary work is done by highly competent people, namely scientists.

• Scientific publishing is one of the very few branches of industry in which the producers are openly criticized by their main customers, the libraries. In fact, in 1998, the Scholarly Publishing and Academic Resources Coalition, or SPARC (www.arl.org/sparc), was founded to support creation of new, inexpensive scientific journals. Currently more than 200 libraries joined SPARC. SPARC recently produced a *Declaring Independence Handbook* [5] explaining to scientific journal editors what options they have if they find their journal is too expensive.

• The market forces in scientific publishing do not play the same role as in other areas of industry. In particular, it is difficult to enforce a price reduction or to move a scientific journal from one publisher to another. A recent laudable example is Elsevier's *Journal of Logic Programming*. The entire editorial board left Elsevier to found a new journal, *Theory and Practice of Logic Programming*, published by the Cambridge University Press at a 60% lower price per page. An account of how this happened can be found in [2].

## For-Profit Versus Non-Profit Publishers

Before we proceed it is important to distinguish between for-profit (commercial) and non-profit publishers. In computer science the first type is represented by Reed Elsevier, Wolters Kluwer,

# Viewpoint

Springer-Verlag, John Wiley, and others. The second type is represented by the academic publishers (for example, Cambridge University Press, Princeton University Press, MIT Press, to name a few)—and publishers associated with the learned societies such as ACM, IEEE, and SIAM.

For a number of years the scientific publishers have increased the subscription prices to scientific

> We were brought up in the tradition of accepting the publisher as a guarantor of quality. In reality the quality is provided by us, the researchers, by means of the peer review process organized by us, for us. But we still need the publisher for its *trademark*.

journals by a margin exceeding the inflation rate; for-profit publishers increased their prices much more. According to the Association of Research Libraries the unit cost of scientific journals increased 169% from 1986 to 1997, while the consumer prices index increased only 46% (see [6]). These price increases led to a dramatic decline in the quality of scientific libraries. The current situation is that conglomerates of large university libraries often have to share the costs to afford access to the portals built by commercial publishers over the journal Web sites.

We entrust publishers with the dissemination of our work but most of the commercial publishers view our publications as just another commercial product. This is not what we meant. Am I wrong here? Would you do voluntary work for Microsoft? How about Coca Cola? After all, we also rely on their products.

If we wish to realize free scientific publishing (FSP), it is clear the commercial publishers won't be interested in cooperation. Instead, we need to work with the non-profit publishers instead.

## Do We Need Publishers at All?
The question is not as strange as it seems, since most of the work for a scientific journal is done nowadays by the researchers themselves. In fact, to run a journal we need:

- volunteers who agree to be editors and referees;
- a volunteer who would create and maintain the home page of the journal;
- a distribution system for making the journal freely available in electronic form;
- financial support for the copy editors, if we wish to adhere to the accepted standards of processing accepted papers; and
- a publisher who would print and bind the articles and take care of the distribution and the subscriptions.

If we ignore for a moment the printed version, the only costs in running a journal are those related to making the articles freely available electronically and the fees for the copy editors. The problem of taking care of grammatical corrections will always remain. Even if we drop the proofreading phase (as a couple of well-known journals have already done, including all FSP journals mentioned in this column) we lower the standards of publishing only a bit.

## FSP Journals under the ACM Auspices
So what are we left with? I argue that if we limit ourselves to an electronic version, we do not need a publisher to publish a high-quality journal. But if I stopped now claiming that we can therefore publish journals for free without publishers, absolutely nothing would change, and most articles would still end up on the Web sites of the journals with a nontrivial price tag attached to them. Why?

Here comes the punch line. Because we were brought up in the tradition of accepting the publisher as a guarantor of quality. In reality the quality is provided by us, the researchers, by means of the peer review process organized by us, for us. But even if we know it, and we know it by now, we still need the publisher for its *trademark*. Even journals published in the void, without any publisher, still need some anchor in reality, be it some university department or even Apt's Wisdom Press. The better the trademark, the better for the journal.

And here is where we need the ACM: to open its digital gates to allow the creation of free journals and to certify them. Each such journal would bear the name of the ACM and would have its Web site on ACM's servers. And with an ACM certificate things would certainly start moving. With the additional support from SPARC these journals could be properly introduced to the libraries and become widely known. Obviously, the ACM Publications Board would have to screen the submitted proposals and monitor the production process to avoid a degeneration of the idea and potential embarrassment. Any respectable group of computer scientists should be allowed to start a free journal under the ACM auspices. Each journal is in fact a microcosmos representing some community of scientists. Each such community has enough enthusiasts and people with some organizational skills and a sense of leadership who would be more than happy to run their own journal, for free, provided it will be certified by the ACM. That is how the ACM *Transactions on Computational Logic* started in 1999 with the inaugural issue published in July 2000. Research in science has always involved a great deal of volunteer work. More than 150 people are listed at ACM's volunteer site (www.acm.org/key_people/volalpha). I am arguing that this good will and dedication can and should be tapped by ACM for a massive creation of FSP journals.

Especially welcome should be rebel groups that leave expensive journals to start ACM alternatives. By doing so the defecting computer scientists will provide a most valuable and honorable service to our community: from that moment on they will pass the accepted articles into the pool of free scientific knowledge.

And to make searching, browsing, and free subscription of such journals possible they could form an overlay of the Computing Research Repository or CoRR (www.arXiv.org/archive/cs/intro.html). CoRR forms a part of the Los Alamos National Laboratory (LANL) archives widely used since 1992 by physicists and mathematicians. The number of connections per day is around 100,000. CoRR opened, in coopera-tion with ACM, in August 1998.

Am I fantasizing? Not at all.

First, please see the *Journal of Artificial Intelligence Research* (JAIR; www.jair.org), founded in 1993 by Steve Minton. In seven years this FSP journal became a highly visible journal in the AI community, indexed by INSPEC, Science Citation Index, and Math-SciNet. In turn, Ulf Rehmann established in 1996 a high-quality FSP mathematical journal *Documenta Mathematica* (www.mathematik.uni-bielefeld.de/documenta). The *Advances in Theoretical and Mathematical Physics* is the first overlay journal. This high-quality FSP journal was founded in 1997 by the 1982 Fields medal winner S.-T. Yau. Its articles consist just of pointers to the LANL archives. Many more successful examples can be cited. ACM journals are one step from becoming FSP journals. In fact, the accepted papers are usually immediately available from the Web site of any of the ACM transactions or from the *Journal of the ACM*. If these papers were kept forever, these journals would become effectively FSP journals.

## Some Questions about Economics

Instead of inventing some economic models, it is much better to rely on public information provided by those who run successful FSP journals. In [4] Minton and Wellman provide a detailed analysis of the economic matters involved in the production of JAIR based on their five years of experience. They find that "the only significant cost involved in producing the journal is the cost of the review and editing process. Thus, the universities and research labs that employ JAIR's editors effectively subsidize the journal by supporting this work." In turn, Louis, Schneider, and Rehmann [3] published an account of the costs of running *Documenta Mathematica* based on their four years of experience. In their detailed analysis they reach a revealing amount of $210 per year ("including hidden costs"). So, not surprisingly, Rehmann wrote to me: "Our journal was never sponsored by anybody. Needless to say, the journal is hosted on my PC, which I have anyway."

In these FSP cases, the administrative work was properly automated. Necessity (here: lack of sponsors

# Viewpoint

and dedication to a noble ideal) was mother of invention. The costs of this work can be easily absorbed by the universities and research labs (for example, by trading them for a subscription of an expensive journal).

A discussion about who is to pay the expenses of the order of $210 per year per journal and which fraction of them should be paid to ACM for FSP journals under the ACM auspices is not worth of the glossy (read: expensive) paper of *Communications*.

## What About the Paper Versions?
Most of us are attached to printed journal versions, both as readers and as writers, and there is no compelling reason to abandon printing provided the price will be affordable to all libraries. It is natural to envisage an evolution of the existing ACM journals toward the following model:

- Accepted articles are kept permanently on journal Web sites
- A search engine is added to facilitate the use of this pool of freely available articles
- The final, proofread versions of the articles continue to enter the ACM Digital Library
- The journals are printed and distributed, perhaps only in one bound volume per year, to lower production costs

A legitimate question is what will happen with the subscriptions to the ACM Digital Library and to the printed versions of existing ACM journals. In my opinion not much. These subscriptions are extremely moderately priced and librarians will more likely drop all subscriptions to commercial journals before discontinuing subscriptions of ACM journals.

Actually, the trend is that more and more scientific institutions provide a subscription to the ACM Digital Library and to the printed versions of ACM journals as a global service to all of their employees, as witnessed by the current 340 institutional subscriptions to it. I doubt the aforementioned model would in any way affect this trend.

The fact that the ACM Digital Library is not free is contrary to the spirit of FSP. But if we want to start our discussion on the creation of FSP journals under the ACM auspices by focusing on this problem, the revolution I am calling for will devour its children in no time.

## Computer Scientists Lag Behind
My call for a massive creation of FSP computer journals is by no means original. For example, the Electronic Society for Social Scientists (www.elsss. org.uk) aims at "the provision of electronic publications of high quality, wide diffusion, and low cost for the direct benefit of the academic community."

This recent initiative is supported among others by all economics departments in the U.K. and some 90 scientists from Belgium, Canada, Germany, Israel, Switzerland, the U.K., and the U.S. The organizers report over 1,000 positive responses. The Public Library of Science project (www.publiclibraryof-science.org) has circulated an open letter, and by mid-March 2001 was signed by 11,244 scientists from 119 countries. The signatories from medicine and the life sciences pledge that "beginning in September 2001, we will publish in, edit or review for, and personally subscribe to, only those scholarly and scientific journals that have agreed to grant unrestricted free distribution rights to any and all original research reports they have published." These noble goals of this project should be ours, as well, with ACM taking the lead. **C**

**KRZYSZTOF R. APT** (k.r.apt@cwi.nl) is a senior researcher at CWI, Amsterdam, and a professor of computer science at the University of Amsterdam.